\documentstyle[12pt,epsf,psfig]{article}
\begin{document} 
\title{\bf  Effects of the Fourth Generation on 
$\Delta M_{B_{d,s}}$ in  $B^0-\bar B^0$ Mixing}
 \author{Wu-Jun Huo\\ {\sl
CCAST (World Lab.), P.O. Box 8730, Beijing $100080$}\\{\sl  and}\\{\sl
Institute of High Energy Physics, Academia Sinica, P.O. Box $918(4)$},\\{\sl 
Beijing $100039$, P.R.  China}}

\date{} 
\maketitle

\begin{abstract}

 We investigate the mass differences $\Delta M_{B_{d,s}}$ in the 
  mixing $B^0_{d,s}-\bar B^0_{d,s}$  with a new up-like quark $t^{\prime}$ in a 
 sequential fourth generation model. We give the basic formulae 
 for $\Delta M_{B_{d,s}}$ in this  model. (1), then, we analysis 
 the final numerical results of $\Delta M_{B_{s}}$, which is the function
  of $m_{t^{\prime}}$ through a fourth generation CKM factor
  $V^{*}_{t^{'}s}V_{t^{'}b}$ constrained by the rare decay
  $B\rightarrow X_s l^+ l^-$ and two new Wilson coefficients
  $S_0(x_{t^{'}})$ and $S_0(x_t,x_{t^{'}})$. We found that one kind of 
  the new results 
   are satisfied with the present experimental low-bound of $\Delta M_{B_s}$.
  (2), we  get the constraint of the fourth generation CKM factor  
  $V^{*}_{t^{'}b}V_{t^{'}d}$, (which is also the function of $t^{'}$),
  from the experimental measurements  of $\Delta M_{B_d}$ and give
  the figs. of $V^{*}_{t^{'}b}V_{t^{'}d}$ to $m_{t^{'}}$ and the general
  CKM factor $V^{*}_{tb}V_{td}$. We also give the numerical results of 
  the Wilson coefficients $S_0(x_{t^{'}})$, $S_0(x_t,x_{t^{'}})$ and
   $\eta_{t^{'}}$, $\eta_{tt^{'}}$ as the function of $m_{t^{'}}$. 
   We also talk about the hierarchy of the fourth generation CKM matrix.
  As one of the directions beyond the SM, $\Delta M_{B_{d,s}}$
  could provide a possible test of the fourth generation or perhaps
   a signal of the new physics.

\end{abstract}
\newpage

\section{Introduction}

 The Standard Model (SM) is a very successful theory of the elementary
particles known today. But it must be incomplete because it has too many 
unpredicted parameters ($ninteen!$) to be put by hand. Most of these 
parameters are in the fermion part of the theory. We don't know the source 
of the quarks and leptons, as well as
how to determinate their mass and number theoretically. We have to get their 
information all from experiment. There is still no successful 
theory which can be
 descripted them with a unified point, even if the Grand Unified 
 Theory\cite{grand} and Supersymmetry\cite{supers}.
Perhaps {\it elementary particles} have substructure ( like $preon$)\cite{substracture}
 and we need to progress more elementary theories. But this is beyond our 
 current experimental level. 
 
 From the point of phenomenology,
for fermions, there is a realistic question is number of the fermions generation or
weather there are other additional quarks or leptons. The present experiments 
can tell us there are only three generation fermions with $light$ neutrinos
which mass are less smaller than $M_Z /2$\cite{Mark} but the experiments don't 
exclude the existence of other additional generation, such as the fourth
generation, with a $heavy$ neutrino, i.e. $m_{\nu_4} \geq M_Z /2$\cite{Berez}.
Many refs. have studied models which extend the fermions part, such as
vector-like quark models\cite{vec-like}, sterile neutrino models\cite{sterile}
and the sequential four generation standard model (SM4)\cite{McKay} which 
we talk in this note. We consider
 a sequential fourth generation non -SUSY model\cite{McKay}, which is added
  an up-like quark $t^{'}$, a down-like quark $b^{'}$, a lepton $\tau^{'}$, 
  and a heavy neutrino $\nu^{'}$ in the SM. The properties of these new 
  fermions are all the same as their corresponding counterparts 
  of other three generations except their masses and CKM mixing, see tab.1,
  
\begin{table}[htb] 
\begin{center} 
\begin{tabular}{|c||c|c|c|c|c|c|c|c|} 
\hline 
& up-like quark & down-like quark & charged lepton &neutral lepton \\ 
\hline 
\hline 
& $u$ & $d$& $e$ & $\nu_{e}$ \\ 
SM fermions& $c$&$s$&$\mu$&$\nu_{\mu}$ \\ 
& $t$&$b$&$\tau$&$\nu_{\tau}$\\
\hline
\hline new
fermions& $t^{'}$&$b^{'}$&$\tau^{'}$&$\nu_{\tau^{'}}$ \\ 
\hline 
\end{tabular}
\end{center}
\caption{The elementary particle spectrum of SM4} 
\end{table}

There are a lot of refs. about the fourth generation. Some refs. devoted to
the mass spectrum of the fourth generation particles\cite{McKay}, as well as 
some discussed the mass bounds of the fourth leptons\cite{lepton}. There are 
many papers talked about other problems about the fourth 
generation\cite{huo1,huo2,fourth}
and the experimental search of the fourth generation paticles\cite{exp1,exp}.
In our previous papers\cite{huo1,huo2}, we investigated the rare $B$ meson
decays with the fourth generation\cite{huo1} and $\epsilon^{'} /\epsilon$
in $K^0$ systems in SM4\cite{huo2}. We got some interesting results, such
as the new effects of the 4th generation particle on the meson decays and 
CP violation. We also got the constraints of the fourth generation CKM matrix
factors, like $V_{t^{'}s}^{*} V_{t^{'}}b$ from 
$B\rightarrow X_s \gamma$\cite{huo1} and $V_{t^{'}s}^{*} V_{t^{'}}d$ from
$\epsilon^{'} /\epsilon$\cite{huo2}. In other words, these rare decays
provided possible test of the fourth generation existence, as well as
CP violation.
In this note, we talk about the mass difference $\Delta M_{B_{d,s}}$ in 
$B^0 -\bar B^0$\cite{BB,ali} with the fourth
generation. We will give the prediction of $\Delta M_{B_s}$ 
in SM4 and the constraints of a new fourth generation CKM matrix factor
$V_{t^{'} b}^{*} V_{t^{'}}d$ from $\Delta M_{B_d}$ in SM4. Like 
refs.{huo1,huo2}, this note can also provide a test of 4th generation.

Particle-antiparticle mixing is responsible for the small mass 
differences between the mass eigenstates of neutral mesons, such
as $\Delta M_K$ in $K_{\rm L} -K_{\rm S}$ mixing and $\Delta M_{B_{d,s}}$
in $B^0 -\bar B^0$ mixing. Being an FCNC process it involves heavy
quarks in loops and consequently it is a perfect testing ground for
heavy flavor physics. For example, $B^0 -\bar B^0$ mixing\cite{bdb}
gave the first indication of a large top quark mass. 
$K_{\rm L} -K_{\rm S}$ mixing is also closely related to the violation of 
CP symmetry which is experimentally known since 1964\cite{cpe}.
They are sensitive measures of the top quark $t$ couplings $V_{ti}(i=d,s,b)$
and of the top quark mass $m_{t}$. The experimental measurements of 
$\Delta B_{d}$ is used to determine the CKM matrix elements $V_{td}$\cite{BB}.
 It  offer an improved determination of the unitarity triangle with 
 the future accurate measurement of $\Delta M_{B_s}$\cite{BB,ali}. 
 For physics beyond the SM, there 
 are a number of studies of the new physics effects in $B_d$ 
 decays\cite{bdn,ali,mssm}. But $B_s$ system has received somewhat less attention 
 from new physics point of view\cite{bsn,ali,mssm}. Experimentally, $\Delta M_{B_d}$
 has been accurately measured,
 $\Delta M_{B_d} =0.473\pm0.016(ps)^{-1}$\cite{ali,expm}.
 But $\Delta M_{B_s}$ has only lower bound, 
 $\Delta M_{B_s} >14.3(ps)^{-1}$\cite{ali,2059,expm}. 
 
 In this note, We want to investigate $\Delta M_{B_{d,s}}$ in $B^0 -\bar B^0$
 mixing in SM4. First, if we add a sequential fourth up-like quark $t^{'}$,
 there produce new prediction of the the mass difference $\Delta M_{B_s}$
 through the new Wilson coefficients and a fourth generation CKM matrix
 factor $V^{*}_{t^{'} s} V_{t^{'} b}$, which constrained for the rare decay
 $B\rightarrow X_s \gamma$ in \cite{huo1}. We found, as like the analysis of 
 $B\rightarrow X_s l^+ l^-$ in \cite{huo1}, our results of the prediction
 of $\Delta M_{B_s}$ in SM4 are quite different from that of SM  and
 can satisfy the lower experimental bound  in one case
 of the values $V^{*}_{t^{'} s} V_{t^{'} b}$ taken. In another case, it is
 almost the same as the SM one. The new effects of the fourth generation 
 show clearly in the first case. Second, we can get the constraint of a
 fourth generation CKM matrix factor, $V^{*}_{t^{'} b} V_{t^{'} d}$ from
 the experimental measurement of $\Delta M_{B_d}$. We get one kind of reasonable
 analytical solutions of $V^{*}_{t^{'} b} V_{t^{'} d}$.  They
  are very small,   $-1.0 \times 10^{-4} \leq V^{*}_{t^{'} b} V_{t^{'} d} 
   \leq 0.5 \times 10^{-4}$. These result don't contradicted the unitarity
   constraints for quark $d,b$\cite{data}. Moreover, This small absolute value,
   $\lambda^{-4} \sim \lambda^{-5}$ order, is in agreement with the hierarchy
 in the CKM matrix elements\cite{wolfenstein,Wakaizumi}. It seems to give the possible test of
 this hierarchy and the existence of the fourth generation\cite{Wakaizumi}. We give
 the analysis of the hierarchy in the four generation CKM matrix.
  
In sec. 2,  we give the basic formulae for the mass difference 
$\Delta M_{B_{d,s}}$ in $B^0 -\bar B^0$ with the sequential fourth 
generation up-like quark $t^{'}$ in SM4 model.
In sec. 3, we give the prediction of mass difference $\Delta M_{B_s}$
in SM4 and the numerical analysis.
 Sec. 4 is devoted to the numerical analysis of one fourth generation
 CKM matrix factor $V^{*}_{t^{'}b}V_{t^{'}d}$ from the experimental
 measurements of the mass difference $\Delta M_{B_d}$ in SM4. We also 
 analyze the hierarchy of the four generation CKM matrix in this section.
Finally,  in sec. 5, we give our conclusion.
  
\section{Basic formulae for $\Delta M_{B_{d,s}}$   with $t^{'}$}

 $B^0_{d,s}-\bar B^0_{d,s}$ mixing proceeds to an excellent approximation only 
 through box diagrams with internal top quark exchanges in SM. In SM, the effective 
 Hamiltonian ${\cal H}_{\rm eff}(\Delta B=2)$ for $B^0_{d,s}-\bar B^0_{d,s}$
 mixing, relevant for scales $\mu_b={\cal O}(m_b)$ is given by\cite{BB}
 \begin{eqnarray}
 {\cal H}_{\rm eff}^{\Delta B=2}=\frac{G_{\rm F}^2}{16\pi^2}{M_W^2}
 (V^{*}_{tb}V_{tq})^2 S_0(x_t) Q(\Delta B=2)+h.c. 
 \end{eqnarray}
where $ Q(\Delta B=2)=(\bar b_\alpha q_\alpha)_{V-A}
  (\bar b_\beta q_\beta)_{V-A }$, with $q=d,s$ for $B^0_{d,s}-\bar B^0_{d,s}$
respectively and $S_0(x_t)$ is the Wilson coefficient which is taken the form
 \cite{BB} 
\begin{eqnarray}
S_0 (x_t)=\frac{4x_t -11x_t^2 +x_t^3}{4(1-x_t )^2}
 -\frac{3}{2}\cdot \frac{x_t^3}{(1-x_t )^3}\cdot\ln{x},
\end{eqnarray}
where $x_t =m_t^2 /M_W^2$.
 The mass differences $\Delta M_{d,s}$ can be expressed 
 in terms of the off-diagonal element in the neutral $B$-meson mass matrix 
 \begin{eqnarray}
 \Delta M_{d,s}&=&2|M^{d,s}_{12}|\\
2m_{B_{d,s}}|M^{d,s}_{12}|  &= &|\langle \bar B_{d,s}^0|{\cal H}_{\rm eff}(\Delta B=2)
               |B_{d,s}^0\rangle . \nonumber
 \end{eqnarray}
 
 Now, we turn to the case of SM4. If we add a fourth sequential fourth generation up-like quark
$t^{\prime}$, the above equations would have some modification.
There exist other box diagrams contributed by $t^{'}$ (see fig. 1), similar to
the leading box diagrams in MSSM\cite{mssm}.

The effective Hamiltonian in Standard Model, eq.(1), chang into the form\cite{eta},
 \begin{eqnarray}
 {\cal H}_{\rm eff}^{\Delta B=2}&=&\frac{G_{\rm F}^2}{16\pi^2}{M_W^2}
 [\eta_t (V^{*}_{tb}V_{tq})^2 S_0(x_t)+ \nonumber \\
 &+ &\eta_{t^{'}} (V^{*}_{t^{'}b}V_{t^{'}q})^2
  S_0(x_t^{'})+\eta_{tt^{'}} (V^{*}_{t^{'}b}V_{t^{'}q})\cdot 
   (V^{*}_{tb}V_{tq}) S_0(x_t,x_t^{'})] 
  Q(\Delta B=2)+h.c. 
 \end{eqnarray}
 The mass differences $\Delta M_{d,s}$ in SM4 can be expressed 
 \begin{eqnarray}
 \Delta M_{d}&=&\frac{G_{\rm F}^2}{6\pi^2}{M_W^2}{m_{B_{d}}}
 (\hat B_{B_{d}}\hat F^2_{B_{d}})
 [\eta_t (V^{*}_{tb}V_{td})^2 S_0(x_t)+ \nonumber\\
 & +&\eta_{t^{'}} (V^{*}_{t^{'}b}V_{t^{'}d})^2
  S_0(x_{t^{'}})+\eta_{tt^{'}} (V^{*}_{t^{'}b}V_{t^{'}d})\cdot
  (V^{*}_{tb}V_{td}) S_0(x_t,x_{t^{'}})]\\
 \Delta M_{s}&=&\frac{G_{\rm F}^2}{6\pi^2}{M_W^2}{m_{B_{s}}}
 (\hat B_{B_{s}}\hat F^2_{B_{s}})
 [\eta_t (V^{*}_{tb}V_{ts})^2 S_0(x_t)+ \nonumber\\
 & +&\eta_{t^{'}} (V^{*}_{t^{'}b}V_{t^{'}s})^2
  S_0(x_{t^{'}})+\eta_{tt^{'}} (V^{*}_{t^{'}b}V_{t^{'}s}) \cdot
  (V^{*}_{tb}V_{ts}) S_0(x_t,x_{t^{'}})]
 \end{eqnarray}
 where $(\hat B_{B_{s}}\hat F^2_{B_{s}})=\xi_s^2 
 \cdot (\hat B_{B_{d}}\hat F^2_{B_{d}})$. 
 The new Wilson  coefficients $S_0(x_{t^{'}})$ 
present the contribution of $t^{\prime}$,
which  like  $S_0(x_{t})$  SM in eq. (5)
except exchanging $t^{\prime}$ quark not $t$ quark.
 $S_0(x_t,x_{t^{'}})$ present the contribution of a mixed $t-t^{\prime}$,
  which is taken the form\cite{sxy}
\begin{eqnarray}
S_0(x,y)&=&x\cdot y [-\frac{1}{y-x}(\frac{1}{4}+\frac{3}{2}\cdot \frac{1}{1-x}
 -\frac{3}{4}\cdot \frac{1}{(1-x)^2}\ln{x}+ \nonumber \\
 &+&(y\leftrightarrow x)-\frac{3}{4}\cdot \frac{1}{(1-x)(1-y)}]
 \end{eqnarray}
 where $x= x_t =m_t^2 /M_W^2$, $y= x_{t^{'}} =m_{t^{'}}^2 /M_W^2$. The 
 numerical results of $S_0 (x_{t^{\prime}})$ and $S_0 ({x_t}, {x_{t^{\prime}}})$
 is shown on the tab. 2.
\begin{table}[htb]
\begin{center}
\begin{tabular}{|c||c|c|c|c|c|c|c|c|c|c|}
\hline
$m_t^{'}$(GeV) &50  &100  & 150 & 200 & 250 &300 &350 &400&450&500   \\
\hline
$S_0 (x_{t^{'}})$ &0.33&1.07&2.03&3.16&4.44&5.87&7.47&9.23&11.15&13.25 \\
$S_0 (x_t ,x_{t^{'}})$ & 0.48&-7.03&-4.94&-5.09& -5.39&-5.87&-5.99&-6.25
&-6.49& -6.72 \\
\hline
\hline
$m_t^{'}$(GeV) &550  &600  & 650 & 700 &750 &800 &850 &900&950&1000   \\
\hline
$S_0 (x_{t^{'}})$ &15.52&17.97&20.60&23.41&26.40&29.57&
32.93 & 36.47 & 40.96 & 44.11  \\
$S_0 (x_t ,x_{t^{'}})$ &-6.92& -7.11&-7.28&-7.44&-7.60&-7.74& -7.87
& -7.99& -8.12 & -8.23 \\
\hline
\end{tabular}
\end{center}
\caption[]{The Wilson coefficients $S_0 (x_{t^{\prime}})$ 
and $S_0 ({x_t}, {x_{t^{\prime}}})$ to  $m_{t^{'}}$}
\end{table}

The short-distance QCD correction  factors 
$\eta_{t^{'}}$ and $\eta_{tt^{'}}$ can be
calculated like $\eta_c$ and $\eta_{ct}$ in the mixing of $K^0 - \bar K^0$,
 which the NLO values are given in refs\cite{BB,eta}, 
relevant for scale not ${\cal O}(\mu_c)$ but ${\cal O}(\mu_b)$. 
In leading-order, $\eta_{t}$ is calculated by
\begin{eqnarray}
\eta^0_t =[ \alpha_s (\mu_t )]^{(6/23)},\ \ \ 
  \alpha_s (\mu_t) =\alpha_s (M_Z) [1+\sum^{\infty}_{n=1}
 (\beta_0 \frac{\alpha_s (M_Z)}{2\pi} {\rm In}\frac{M_Z}{\mu_t})^n],
\end{eqnarray}
with its numerical value in tab. 3. The formulae of factor $\eta_{t^{'}}$ is
similar to the above equation except for exchanging $t$ by $t^{'}$.In
 ref.\cite{Wakaizumi}, to leading order, $\eta_{tt^{'}}$ was taken as 
\begin{eqnarray}
\eta^0_t{'} =[ \alpha_s (\mu_t )]^{(6/23)}
[ \frac{\alpha_s (\mu_{b^{'}} )}{\alpha_s (\mu_t )}]^{(6/21)}
[ \frac{\alpha_s (\mu_{t^{'}} )}{\alpha_s (\mu_{b^{'}} )}]^{(6/19)},
\end{eqnarray}
with value 0.58 for the same as for $\eta^K_{tt^{'}}$ in $K^0 -\bar K^0$.
For simplicity, we take $\eta_{tt^{'}} =\eta_{t^{'}}$. We give the 
numerical results in tab.4.

\begin{table}[htb]
\begin{center}
\begin{tabular}{|c||c|c|c|c|c|c|c|c|c|c|}
\hline
$m_t^{'}$(GeV) &50  &100  & 150 & 200 & 250 &300 &350 &400&450&500   \\
\hline
$\eta_{t^{'}}$ &0.968&0.556&0.499&0.472&0.455&0.443&0.433&0.426&0.420&
0.416 \\
\hline
\hline
$m_t^{'}$(GeV) &550  &600  & 650 & 700 &750 &800 &850 &900&950&1000   \\
\hline
$\eta_{t^{'}}$ &0.412&0.408&0.405&0.401&0.399&0.396&
0.395 & 0.393 &0.391 &0.389  \\
\hline
\end{tabular}
\end{center}
\caption[]{The short-distance QCD factors  $\eta_{t^{'}}$,
$\eta_{t{t^{'}}}(=\eta_{t^{'}})$ to  $m_{t^{'}}$}
\end{table}

In the last of this section, we give other input parameters necessary
in this note. (See the following tab.).

\begin{table}[htb]
\begin{center}
\begin{tabular}{ |c| c| c| c | }
\hline
 $\overline m_c(m_c(pole))$ & $1.25\pm0.05$GeV & $M_W$&$80.2$GeV\\
 $\overline m_t(m_t(pole))$ & $175$GeV &$\hat F_{B_{d}}\sqrt{\hat B_{B_{d}}})$
 & $215\pm40$MeV\\
 $\Delta M_{B_d}$& $(0.473\pm0.016)(ps)^{-1}$ &$\xi_s$ & $1.14\pm0.06$  \\
 $\Delta M_{B_s}$& $>14.3(ps)^{-1}$ & $G_{\rm F}$&$1.166\times10^{-5}$GeV${^{-2}}$   \\
 \hline
\end{tabular}
\end{center}
\caption{Neumerical values of the input parameters\cite{ali}.}
\end{table}

\section{Prediction of $\Delta M_{B_s}$ with $t^{'}$}

Experimentally, the mass difference $\Delta M_{B_s}$ of the 
$B^0 -\bar B^0$ mixing is unclear. It has only low bound,
$\Delta M^{\rm exp}_{B_s} > 14.3 (ps)^{-1}$\cite{ali,2059}.
We have given the calculation formula of $\Delta M_{B_s}$ in eq. (6)
and the numerical results of Wilson coefficients $S_0$ and QCD
correction coefficients $\eta$. Now, if we  constrain the 
fourth generation CKM factor $V^{*}_{t^{'}b}V_{t^{'}s}$, we can predict
$\Delta M_{B_s}$ in our four generations model. 
Fortunately, from our previous paper\cite{huo1}, 
we have given the constraints of $V^{*}_{t^{'}b}V_{t^{'}s}$
from experimental measurements of $B \to X_s \gamma$. Here,  we give only 
the basic scheme and the final numerical results.

 The leading logarithmic calculations can be summarized in a
  compact form  as follows \cite{BB}:
    \begin{equation}\label{main}
      R_{{\rm quark}} =\frac{Br(B \to X_s \gamma)}
       {Br(B \to X_c e \bar{\nu}_e)}=
     \frac{|V_{ts}^* V_{tb}^{}|^2}{|V_{cb}|^2} 
     \frac{6 \alpha}{\pi f(z)} |C^{\rm eff}_{7}(\mu_b)|^2\,.
    \end{equation}
   In the case of four generation there is an additional contribution to $B\rightarrow X_s\gamma$ 
from the virtual exchange of the fourth generation up quark $t^{'}$. 
The Wilson coefficients of the dipole operators are given by
   \begin{equation}
      C^{\rm eff}_{7,8}(\mu_b)=C^{\rm (SM)\rm eff}_{7,8}(\mu_b)
      +\frac{V^{*}_{t^{'}s}V_{t^{'}b}}{V^{*}_{ts}V_{tb}}C^{(4)
      {\rm eff}}_{7,8}(\mu_b),
    \end{equation}
 where $C^{(4){\rm eff}}_{7,8}(\mu_b)$ present the contributions of 
 $t^{'}$ to the Wilson coefficients, and 
$V^{*}_{t^{'}s} V_{t^{'}b}$ are the fourth generation CKM matrix factor 
which we need now.
  With these Wilson coefficients and the experiment results of the decays of
 $B\rightarrow X_{s}\gamma$ and $Br(B \to X_c e \bar{\nu}_e)$ \cite{al,data}, 
 we obtain  the results of the fourth generation CKM factor 
 $V^{*}_{t^{'}s}V_{t^{'}b}$. 
 There exist two cases, a positive factor and a negative one: 
    \begin{eqnarray}
      V^{*}_{t^{'}s}V_{t^{'}b}^{(+)} &=& [C^{(0){\rm eff}}_{7}(\mu_b)
        - C^{\rm (SM)\rm eff}_{7}(\mu_b)]
       \frac{V_{ts}^* V_{tb}}{C^{(4){\rm eff}}_{7}(\mu_b)} \nonumber \\
        &=&
      [\sqrt{\frac{R_{\rm quark}|V_{cb}|^2\pi f(z)}{
        |V_{ts}^* V_{tb}|^2 6 \alpha}}-C^{\rm (SM)\rm eff}_{7}(\mu_b)]
        \frac{V_{ts}^* V_{tb}}{C^{(4){\rm eff}}_{7}(\mu_b)}
    \end{eqnarray}
    \begin{equation}
    V^{*}_{t^{'}s}V_{t^{'}b}^{(-)}=[-\sqrt{\frac{R_{\rm quark}|V_{cb}|^2\pi 
    f(z)}{|V_{ts}^* V_{tb}|^2 6 \alpha}}-C^{\rm (SM)\rm eff}_{7}(\mu_b)]
    \frac{V_{ts}^* V_{tb}}{C^{(4){\rm eff}}_{7}(\mu_b)}
   \end{equation}
as in table. 3. With these values, we can give the prediction of
 $\Delta M_{B_s}$ in SM4 by the figs. 2. It is very interesting
 that the fanal analytic result is same as that in decay of 
 $B\rightarrow X_s l^{+} l^{-}$\cite{huo1}.

\begin{table}[htb]
\begin{center}
\begin{tabular}{|c||c|c|c|c|c|c|c|}
\hline
 $m_{t^{'}}$(Gev) & 50 & 100 & 150 & 200 &250 &300 &350 \\
\hline
$V^{*}_{t^{'}s}V_{t^{'}b}^{(+)}/ 10^{-2}$ & $-11.591$ & $-9.259$&$-8.126$
& $-7.501$ & $-7.116$ & $-6.861$ & $-6.580$  \\
$V^{*}_{t^{'}s}V_{t^{'}b}^{(-)}/ 10^{-3}$ & $3.564$ & $2.850$ & $2.502$
& $2.309$  & $2.191$  & $2.113$  & $2.205$  \\
\hline
\hline
 $m_{t^{'}}$(Gev) & 400 & 500 & 600 & 700 &800 &900 &1000  \\
\hline
$V^{*}_{t^{'}s}V_{t^{'}b}^{(+)}/ 10^{-2}$ & $-6.548$ &  $-6.369$ &$-6.255$ 
& $-6.178$ & $-6.123$ & $-6.082$ & $-6.051$ \\
$V^{*}_{t^{'}s}V_{t^{'}b}^{(-)}/ 10^{-3}$ & $2.016$ & $1.961$&
$1.926$ & $1.902$ & $1.885$ & $1.872$ & $1.863$ \\
\hline
\end{tabular}
\end{center}
\caption{The values of $V^{*}_{t^{'}s}\cdot V_{t^{'}b}$ due to masses of 
$t^{'}$ for $Br(B\rightarrow X_{s}\gamma)=2.66\times 10^{-4}$}
\end{table}

The mass difference $\Delta M_{B_s}$ in the two cases of 
$V^{*}_{t^{'}s}V_{t^{'}b}$ are shown in the figs. 2(a) and 2(b)
respectively. In the first case, which the value of 
$V^{*}_{t^{'}s}V_{t^{'}b}$ takes positive, i.e. 
$(V^{*}_{t^{'}s}V_{t^{'}b}^{-})$,
the curve of $\Delta M_{B_s}$ to $m_{t^{'}}$ is almost overlap with that of SM.
That is, the results in SM4 are the same as that in SM, except a peak in the 
curve when $m_{t^{'}}$ takes value about 170GeV. The reason is not because
there is new prediction deviation from the SM but only because
there is a term of $(x-y)$ in denominator of the formulae of eq. (7). 
 In this case, it does not show the new effects of $t^{'}$.  The mass 
 difference $\Delta M_{B_s}$ is nathless unclear. 
 Also, we can not obtain the information of existence 
of the fourth generation from $\Delta M_{B_s}$, although we can not 
exclude them either.  This is because, from
tab. 5, the values of $V^{*}_{t^{'}s}V_{t^{'}b}^{(-)}$ are positive. But they 
are of order $10^{-3}$ and is very small.  The values of $V^{*}_{ts}V_{tb}$ 
are about ten times larger than them ($V^{*}_{ts}= 0.038$, $V_{tb}=0.9995$, 
see ref. \cite{data} ). Furthermore, the last two terms about $m_{t^{'}}$
in eq. (6) are approximately same order.  The contribution of them counteract
each other. 

But in the second case, when the values of $V^{*}_{t^{'}s}V_{t^{'}b}$  are
negative, i.e. $(V^{*}_{t^{'}s}V_{t^{'}b}^{(-)})$. The curve of $\Delta M_{B_s}$
is quite different from that of the SM. This can be clearly seen from fig. 2(b).
  The enhancement of $\Delta M_{B_s}$ increases rapidly with  increasing of 
  $t^{'}$ quark  mass.  In this case, the fourth generation effects 
 are shown clearly. The reason is that $V^{*}_{t^{'}s}V_{t^{'}b}^{(+)}$ 
 is 2-3 times larger than  $V^{*}_{ts}\dot V_{tb}$ so that the last two terms 
 about $m_{t^{'}}$ in eq. (6)  becomes important and it depends on the $t^{'}$ 
 mass strongly.  Thus, the effect of the fourth generation is significant. 
 Meanwhile, The prediction of $\Delta M_{B_s}$ in SM4 can satisfy the 
 experimental low bound of $\Delta M_{B_s} \geq 14.3(ps^{-1})$. So, the 
 sequential fourth generation model could be one of the ways of searching
 new physics about $\Delta M_{B_s}$. If $V^{*}_{t^{'}s}V_{t^{'}b}$ choose
 this case, the mass difference $\Delta M_{B_s}$ in $B^0 -\bar B^0$ mixing 
 could be a good probe to the existence of the fourth generation.

\section{Constrains of the fourth generation CKM factor 
$V^{*}_{t^{'}b}V_{t^{'}d}$
from experimental measurements of $\Delta M_{B_d}$}

Unlike $\Delta M_{B_s}$, the mass difference $\Delta M_{B_d}$ of 
$B^0_d -\bar B^0_d$ mixing is experimental clear,
$\Delta M^{\rm exp}_{B_d} =0.473\pm 0.016 (ps)^{-1}$\cite{ali}. 
We can get the constraints of the fourth generation CKM factor
$V^{*}_{t^{'}b}V_{t^{'}d}$ from the present experimental value
of $\Delta M_{B_d}$. 

We change the form of eq. (5) as a quadratic equation about
$V^{*}_{t^{'}b}V_{t^{'}d}$. By solving it , we can get two analytical solution
$ V^{*}_{t^{'}d}V_{t^{'}b}^{\rm (1)}$ (absolute value is the large one) and 
 $V^{*}_{t^{'}d}V_{t^{'}b}^{\rm (2)}$ (absolute value is the small one),
 just like the other 4th generation CKM matrix factor 
 $V^{*}_{t^{'}s}V_{t^{'}b}^{(\pm)}$\cite{huo2} in last section.

  However, experimentally, it is not accurate for the measurement of CKM matrix 
  element $V_{td}$\cite{BB,data}. So, we have to search other 
 ways to solve this difficulty. Fortunately,  the  CKM unitarity 
 triangle\cite{Ali}, i.e. the graphic representation of the unitarity relation for 
 $d,b$ quarks,
 which come from the orthogonality condition on the first and
 third row of $V_{\rm CKM}$,
 \begin{eqnarray}
 V_{ud} V^{*}_{ub} + V_{cd} V^{*}_{cb} + V_{td} V^{*}_{tb} =0, 
 \end{eqnarray}
  can be conveniently depicted as a triangle relation in the complex
 plane, as shown in the fig. 3(a).

  From the above equation, we can  give the constraints of 
 $V_{td} V^{*}_{tb}$\cite{9905397},
 \begin{equation}
 0.005 \leq |V_{td} V^{*}_{tb} | \leq 0.013
 \end{equation}
Then, we give the final results as shown in the  figs. 4(a) and 4(b).    

We must announce that figs. 4 only show the curves with 
$V^{*}_{t^{'}d}V_{t^{'}b}^{\rm (2)}$ (absolute value is the small one) firstly.
 Because
the absolute value of $V^{*}_{t^{'}d}V_{t^{'}b}^{\rm (1)}$  is generally larger
than 1. This is contradict to the unitarity of CKM matrix. So, we don't
think about this solution. From the figs. 4, we found all curves are in the range
from $-1\times 10^{-4}$ to $0.5 \times 10^{-4}$ when we considering 
the constraint of $V_{td} V^{*}_{tb}$. That is to say, the absolute value
of $V^{*}_{t^{'}d}V_{t^{'}b}$ is about $\sim 10^{-4}$ order. This is a
very interesting result. 

First, these CKM matrix elements obey unitarity constraints. 
With the fourth generation quark $t^{'}$, eq. (12) change to ,
\begin{equation}
 V_{ud}^{*}V_{ub}+V_{cd}^{*}V_{cb}+V_{td}^{*}V_{tb}
+V_{t^{'}d}^{*}V_{t^{'}b}=0. 
\end{equation} 
This a quadrilatral, (see fig. 3(b)). 
We take the average values of the SM CKM matrix elements from Ref. \cite{data}.
The sum of the first three terms in eq. (12) is about $\sim 10^{-2}$ order.
If we take the value of $V^{*}_{t^{'}s}V_{t^{'}b}^{\rm (2)}$ the result of the 
left of (14) is better and more close to $0$ than that in SM, when
 $V^{*}_{t^{'}s}V_{t^{'}b}^{(2)}$ takes negative values. Even if 
 $V^{*}_{t^{'}s}V_{t^{'}b}^{(2)}$ takes positive values, the sum of (16) 
 would change very little because the values of 
 $V^{*}_{t^{'}d}V_{t^{'}b}^{\rm (2)}$ are about $10^{-4}$ order, two orders 
 smaller than the sum of the first three ones in the left of (14).
Considering that the data of CKM matrix is not very accurate, we can 
get the error range of the sum of these first three terms.  It is  much larger
than $V^{*}_{t^{'}d}V_{t^{'}b}^{\rm (2)}$. Thus,in the case 
 the values of  $V^{*}_{t^{'}d}V_{t^{'}b}$ satisfy the CKM matrix unitarity 
 constraints.

Second, this small order of $V^{*}_{t^{'}d}V_{t^{'}b}$ doesn't contradict to
the hierarchy of the CKM matrix elements or the quarks mixing 
angles\cite{Wakaizumi,hierarchy}. Moreover, 
it seem to prove the hierarchy. The hierarchy in the quarks mixing angles is 
clearly presented in the Wolfenstein parameterization\cite{wolfenstein} of the 
CKM matrix. Let's see CKM matrix firstly,
\begin{equation}
V_{\rm CKM} = \left (
\begin{array}{lcrr}
V_{ud} & V_{us} & V_{ub} & \cdots\\
V_{cd} & V_{cs} & V_{cb} & \cdots\\
V_{td} & V_{ts} & V_{tb} & \cdots\\
\vdots& \vdots& \vdots &\ddots\\
\end{array} \right )
\sim \left (
\begin{array}{lcrr}
1 & \lambda & \lambda^3 & \cdots\\
-\lambda& 1 & \lambda^2 & \cdots\\
\lambda^3 & -\lambda^2 & 1&\cdots\\
\vdots& \vdots& \vdots &\ddots\\
\end{array} \right )
\end{equation}  
with $\lambda=\sin^2 \theta=0.23$. Now, the hierarchy can be expressed in
powers of $\lambda$. We found, the magnitudes of the mixing angles are about 1 among
the $same$ generations, $V_{ud}$, $V_{cs}$ and $V_{tb}$. For different generations,
the magnitudes are about $\lambda$ order between $1st$ and $2nd$ generation,
$V_{us}$ and $V_{cd}$, as well as about $\lambda^2$ order 
between $2nd$ and $3rd$ generation, $V_{cb}$ and $V_{ts}$. The magnitudes 
are about $\lambda^3$ order between the $1st$ and $third$ generation,
$V_{ub}$ and $V_{td}$. Then, there should be an interesting problem: If the fourth
generation quarks exist, how to choose the  order do the magnitude of the mixing
 angles concern the fourth generation quarks? Because there is not direct experimental
 measurement of the fourth generation quark mixing angles, one have to look for other
 indirect methods to solve the problem.
Many refs. have already talked about these additional CKM mixing 
angles\cite{vec-like,sterile,McKay,exp}, like the 
vector-like quark models\cite{vec-like}, the four neutrinos 
models\cite{sterile} and 
the sequential four generations models\cite{McKay}. For simple,
 we give a guess for the magnitude of the fourth generation mixing angles. Similar to
 the general CKM matrix elements magnitude order, the fourth generation
 ones are about $\lambda^4 \sim \lambda^5$ order between the $1st$ and $4th$ 
 generation, such as $V_{t^{'} d}$, as well as $\lambda^2 \sim \lambda^3$ between
 the $2nd$ and $4th$ generation, such as $V_{t^{'} s}$. For the mixing between 
 the $3rd$ and $4th$ generation quarks, such as $V_{t^{'} b}$, we take the magnitude 
 as 1 because the mass of the fourth generation quark $t^{'}$ is the same order,
 $10^2$, as the top quark $t$. So $V_{t^{'} b}$ should take the order of $V_{tb}$.
 Then, the magnitude order of the fourth generation CKM factor 
 $V^{*}_{t^{'}d}V_{t^{'}b}$ is about $\lambda^4 \sim \lambda^5$, i.e. 
 $< \lambda^4$. From figs. 4, we found that
 the  numerical results, $V^{*}_{t^{'}d}V_{t^{'}b}^{(2)}$, satisfy this guess.
 
 At last, the factor $V^{*}_{t^{'}d}V_{t^{'}b}$ constrained from 
 $\Delta M_{B_d}$ does not contradict to the CKM matrix texture. Moreover, it
 seem to support the existence of the fourth generation.

\section{Conclusion}

In this note, we have investigated the mass differences $\Delta M_{B_{d,s}}$ in the 
  mixing $B^0_{d,s}-\bar B^0_{d,s}$  with a new up-like quark $t^{\prime}$ in a 
 sequential fourth generation model. We give the basic formulae 
 for $\Delta M_{B_{d,s}}$ in this  model and calculated the new Wilson coefficients
 $S_0 (m_{t^{'}} /m_W)$ and $S_0 (m_{t^{'}} /m_W)$ in the 
  effective $\Delta S=2$ Hamiltonian  We also calculated the short
 distance QCD factors $\eta_{t^{'}}$ and $\eta_{tt^{'}}$.  With these values, first, we
  analysis  the final numerical results of $\Delta M_{B_{s}}$, which is the function
  of $m_{t^{\prime}}$ through a fourth generation CKM factor
  $V^{*}_{t^{'}s}V_{t^{'}b}$. This factor was constrained from the rare decay
  $B\rightarrow X_s \gamma$ and had two kinds of numerical results.
   We found the new results in SM4 were almost same as in SM with one case 
   of $V^{*}_{t^{'}s}V_{t^{'}b}$ numerical values and  satisfied with the present
   experimental low-bound of $\Delta M_{B_s}$ in other case. We gave the figs of 
   $\Delta M_{B_s}$ to $m_{t^{'}}$ and the numerical analysis.
  Second, we investigated the mass difference $\Delta M_{B_d}$ to get the constraint of 
  the other fourth generation CKM factor  $V^{*}_{t^{'}b}V_{t^{'}d}$,
   (which is the function of $t^{'}$),
  from the experimental measurements  of $\Delta M_{B_d}$. We first got the constraints 
  of the general CKM factor $V^{*}_{tb}V_{td}$ from  the  CKM unitarity 
 triangle. Then we give the figs. of $V^{*}_{t^{'}b}V_{t^{'}d}$ to the mass of $t^{'}$ 
 and to the general factor $V^{*}_{td}V_{tb}$ and the numerical analysis too.
  We also talked about the texture of the fourth generation CKM matrix.
  
 The fourth generation quark $t^{'}$ will give obviously new effects on the
 mass difference $\Delta M_{B_{d,s}}$ if it really exists. At least, the present 
 experimental statue of  $\Delta M_{B_{d,s}}$ could not exclude the space of the 
 fourth generation, Furthmore, the progress of theoretical 
  calculation and experimental measurement  $\Delta M_{B_{d,s}}$
 could provide the strong test of the existence of the fourth generation.
  In other words, as one of the directions beyond the SM, $\Delta M_{B_{d,s}}$
  could provide a possible test of the fourth generation or perhaps
   a signal of the new physics.

\section*{Acknowledgments}
This research was supported in part by the National Nature Science
Foundation of China. I am grateful to prof. C.S. Huang and Prof. Y.L. Wu
for useful discussions and valuable modification andcomments on the manusript.

\newpage
\begin{figure}
\epsfxsize=20cm
\epsfysize=18cm
\centerline{
\epsffile{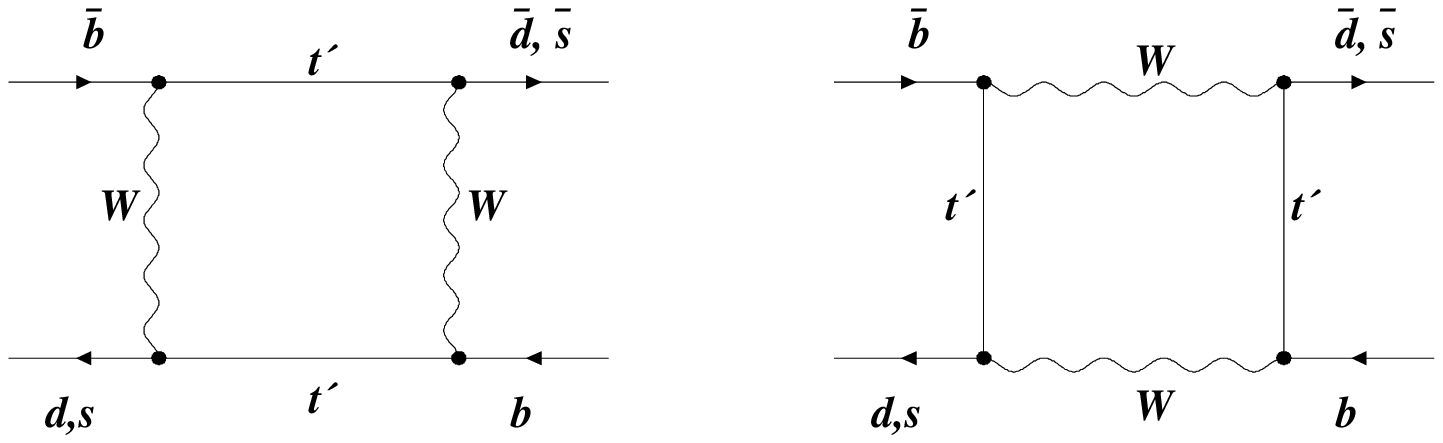}}
\vskip -10.0cm
\caption{The Additional Box Diagrams to $B^0_{d,s} -\bar B^0_{d,s}$
with the fourth up-like quark $t^{'}$.}
\end{figure}

\begin{figure}
\epsfxsize=10cm
\epsfysize=10cm
\centerline{
\epsffile{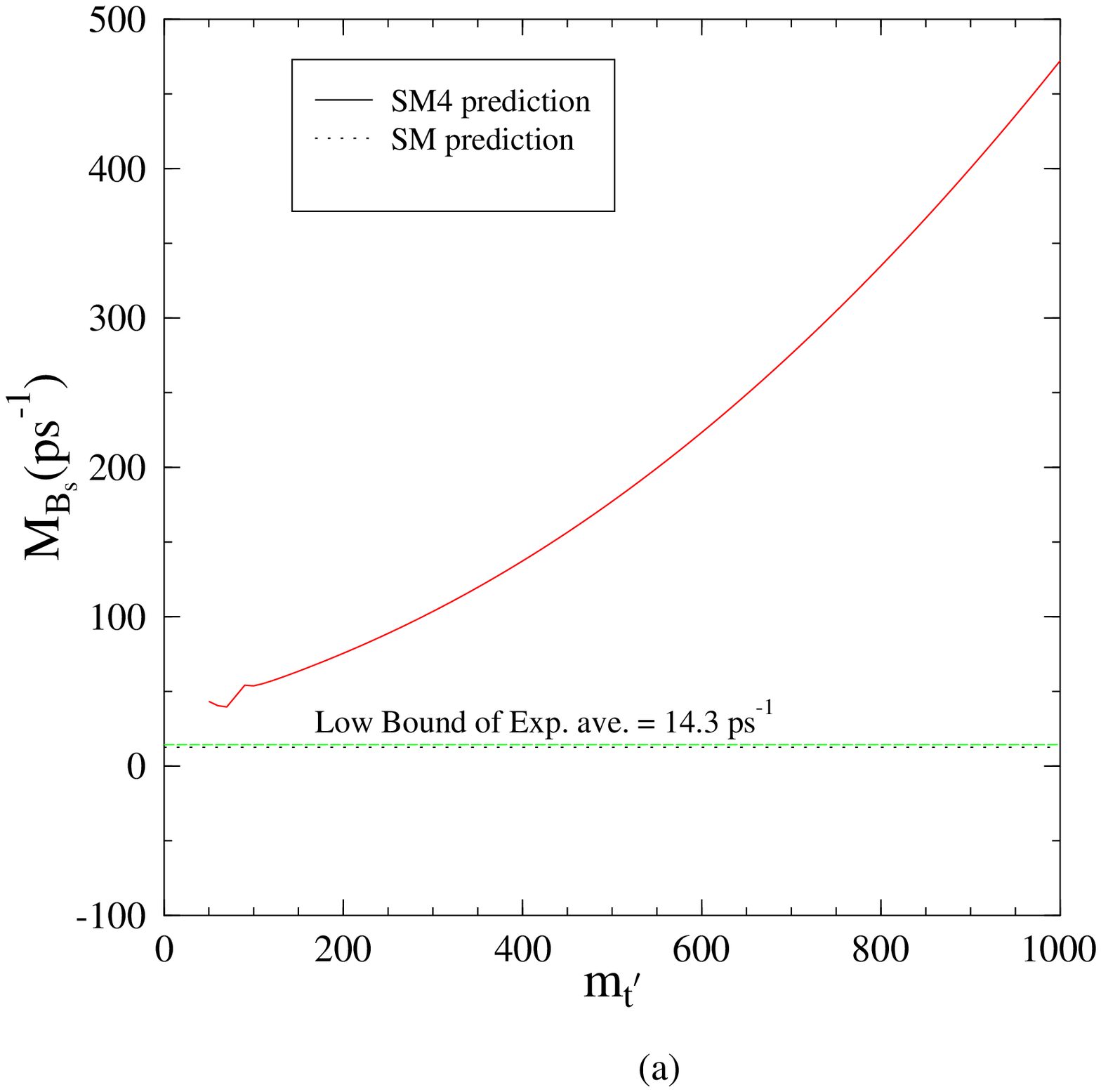}}
\vskip 0cm
\end{figure}

\begin{figure}
\epsfxsize=10cm
\epsfysize=10cm
\centerline{
\epsffile{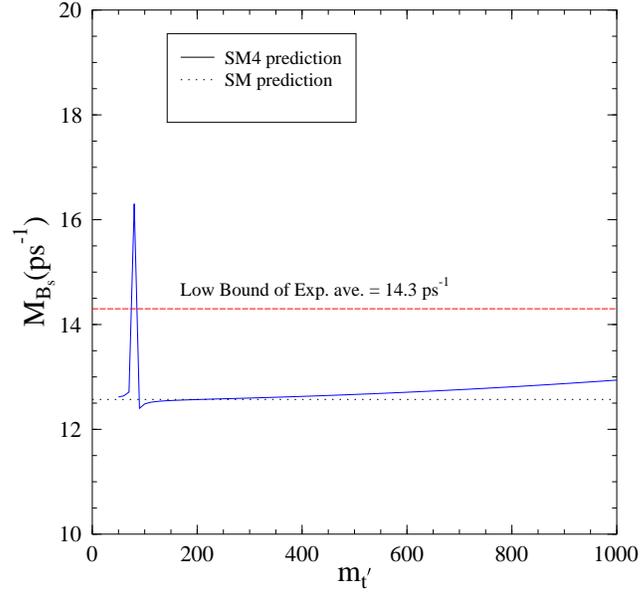}}
\vskip 0cm
\caption{Prediction of $\Delta M_{B_s}$ to $m_{t^{'}}$ in SM4 when
$V^*_{t^{\prime}s} V_{t^{\prime}b}$ takes (a) positive  and (b) negative  values.}
\end{figure}

 \begin{figure}
 \vskip 2cm
\epsfxsize=15cm
\epsfysize=12cm
\centerline{
\epsffile{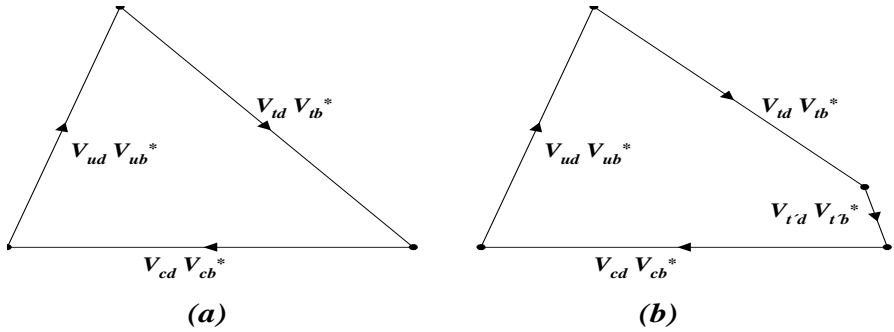}}
\vskip -5cm
\caption{(a) Unitarity triangle for $d,b$ quarks in  Standard Model and 
(b) Unitarity quadrilatral in Sequential 4th Generation Model.}
\end{figure}

\begin{figure}
\epsfxsize=10cm
\epsfysize=10cm
\centerline{
\epsffile{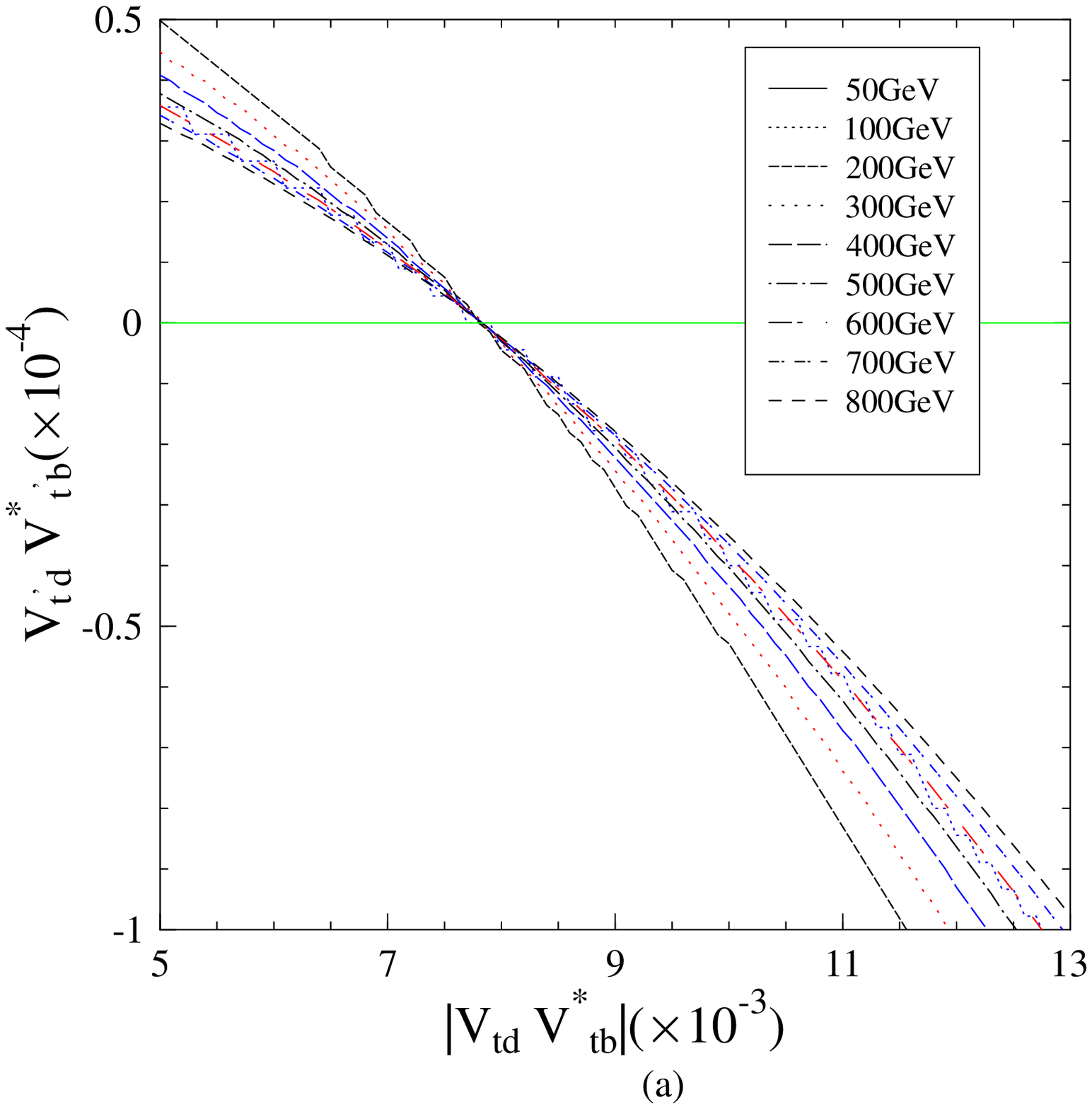}}
\vskip 0cm
\end{figure}

\begin{figure}
\epsfxsize=10cm
\epsfysize=10cm
\centerline{
\epsffile{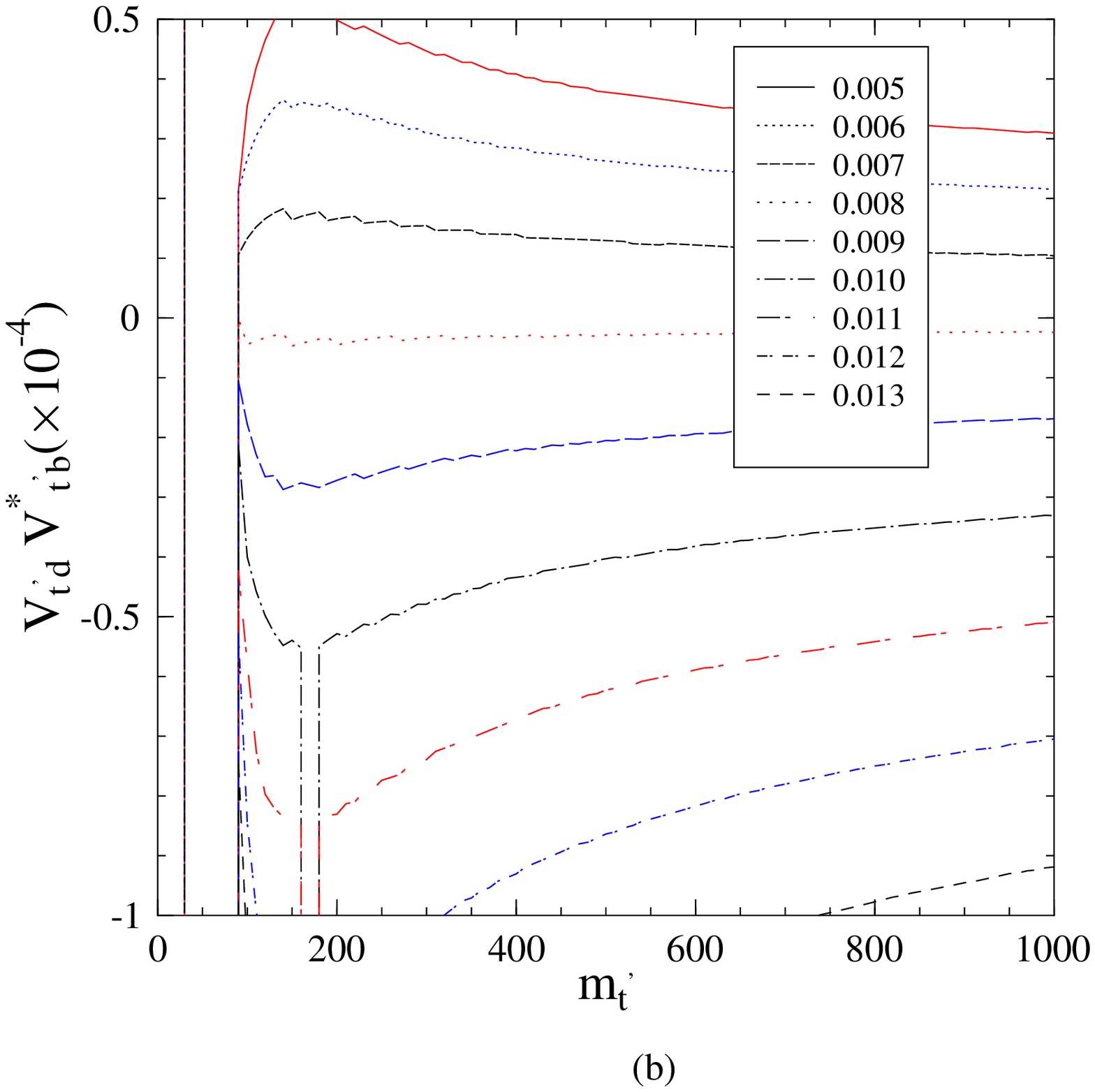}}
\vskip 0cm
\caption{Constraint of the 4th generation CKM factor
$V^{*}_{t^{'}d}V_{t^{'}b}$ to (a) $|V_{td} V^{*}_{tb} |$ with 
 $m_{t^{'}}$ range from 50GeV to 800GeV, (b) to
  $m_{t^{'}}$ with $|V_{td} V^{*}_{tb}|$
range from 0.005 to 0.013.}
\end{figure}

\begin{thebibliography}{99}
\bibitem{grand}P. Langacker, Phys. Rep. {\bf 72}, No. 4, (1981) 185.
\bibitem{supers}M.F. Sohnius, Phys. Rep. {\bf 128}, No. 2\&3 (1985) 39.
\bibitem{substracture}L. Lyons, Prog. Part. Nucl. Phys. {\bf 10} (1983) 227;
C. Heuson, hep-ph/9904493; and references therein.
\bibitem{Mark} G.S.  Abrams et al., Mark II Collab., Phys.  Rev. Lett.  {\bf 63}
(1989) 2173; B.  Advera et al.,
 L3 Collab.,  Phys.  Lett.  B {\bf 231} (1989) 509;I.  Decamp et al.,
 OPAL Collab.,  {\it ibid.}, {\bf 231} (1989) 519; M.Z.  Akrawy et al.,
 DELPHI Collab., {\it ibid.}, {\bf 231} (1989) 539;
 C.Caso et al., (Particle Data Group), Eur.  Phys.  J.C {\bf 3} (1998) 1.  
\bibitem{Berez}Z.  Berezhiani and E, Nardi, Phys.  Rev.  D {\bf 52} (1995) 3087; 
 C.T.  Hill, E.A.  Paschos, Phys.  Lett.  B {\bf 241} (1990) 96.  
\bibitem{vec-like}Y. Nir and D. Silverman, Phys. Rev. {\bf D42} (1990) 1477;
  W-S, Choong and D. Silverman, Phys. Rev. {\bf D49} (1994) 2322; L.T. Handoko, Hep-ph/9708447.
\bibitem{sterile}V. Barger, Y.B. Dai, K. Whisnant and B.L. Young, Hep-ph/9901380;
R.N. Mohapatra, hep-ph/9702229; S. Mohanty, D.P. Roy and U. Sarkar, hep-ph/9810309;
S.C. Gibbons,{\it et al}., Phys. lett. {\bf B430} (1998) 296;
V. Barger, K. Whisnant and T.J. Weiler, Phys. lett. {\bf B427}, (1998) 97;
V. Barger, S. Pakvasa, T.J. Weiler and K. Whisnant, Phys. Rev. {\bf D58} (1998) 093016.
\bibitem{McKay}J.F.  Gunion, Douglas W. McKay, H.  Pois, Phys.  Lett.  B {\bf 334} (1994) 339;
 Phys.  Rev.  D {\bf 51} (1995) 201.
\bibitem{lepton}J. Swain, L. Taylor, hep-ph/9712383;
 R.N.  Mohapatra, X.  Zhang, hep-ph/9301286;
 V. Novikov, hep-ph/9606318.
 \bibitem{huo1} C.S. Huang, W. J. Huo and Y.L. Wu, 
 Mod. Phys. Lett. {\bf A14} (1999)2453.
\bibitem{huo2}C.S. Huang, W. J. Huo and Y.L. Wu,  hep-ph/0005227.

 \bibitem{fourth}L. L. Smith, D. Jain, hep-ph/9501294; 
 K.C. Chou, Y.L. wu, and Y.B. Xie, Chinese Phys.Lett. {\bf 1} (1984) 2.
 A. Datta, hep-ph/9411435;
 T. Yoshikawa, Prog. Theor. Phys. {\bf 96} (1996) 269;
 D.  Grosser,Phys.  Lett.  B{\bf 86} (1979) 301;
 C.D.  Froggatt, H.B.  Nielsen, D.J.  Smith, Z.  Phys. C {\bf 73} (1997) 333;
 S. Dimopoulos, Phys. Lett. B {\bf 129} (1983) 417.
 \bibitem{exp1} LEP1.5 Collab., J. Nachtman, hep-ex/960615.
\bibitem{exp}${\rm D\O}$ Collab., S. Abachi et al., Phys. Rev. Lett. {\bf 78} (1997) 3818.
\bibitem{BB}A.J. Buras, hep-ph/9806471.
\bibitem{ali}A. Ali and D. London, hep-ph/0002167.
\bibitem{bdb}H. Alberecht, {\it et. al.}, Phys. Lett. {\bf B192} (1987) 245;
M. Artuso, {\it et. al}, Phys. Rev. Lett. {\bf 62} (1989) 2233.
\bibitem{cpe}J.H. Christenson, J.W. Cronin, V.L. Fitch and R. Turlay,
Phys. Rev. Lett. {\bf 13} (1964) 128. 
\bibitem{bdn}Y. Grossman and M. Worah, Phys. Lett. {\bf B395}, (1997) 241; 
M.Worah, hep-ph/9711265; N.G. Deshpande, B. Dutta and S. Oh, Phys. Rev. Lett.
{\bf 77}, (1996) 4499; M. Ciuchni, {\it et. al.}, Phys. Rev. Lett. {\bf 79},
(1997) 978; D. London and A. Soni, Phys. Lett. {\bf B407}, (1997) 61;
A. Abd El-Hady and G. Valencia, Phys. Lett. {\bf B414}, (1997) 173; J. P. Silva
and L. Wolfenstein, Phys. Rev. {\bf D55}, (1997) 5331; A.I. Sanda and Z.Z. Xing,
Phys. Rev. {\bf D56} (1997) 6866; S.A. Abel, W.N. Cottingham and I.B.
Whittingham, Phys. Rev. {\bf D58}, (1998) 073006.
\bibitem{mssm}I. Hinchliffe and N. Kersting, hep-ph/0003090.
\bibitem{bsn}G. Barenboim, J. Bernabeu, J. Matias and M. Raidal, hep-ph/9901265;
Y. Grossman, Phys. Lett. {\bf B380} (1996) 99; Z.Z. Xing, Eur. Phys. J. {\bf C4}
(1998) 283; Y. Grossman, Y. Nir and R. Rattazzi, hep-ph/9701231.
\bibitem{expm}G. Blaylock, http://www.cern.ch/LEPBOSC/.
\bibitem{2059}S. Willocq, hep-ex/0002059.
\bibitem{data}C.Caso et al., (Particle Data Group), Eur. Phys. J. C{\bf 3} (1998) 1.
\bibitem{wolfenstein}L. Wolfenstein, Phys. Rev. Lett. {\bf 51} (1983) 1945.
\bibitem{Wakaizumi}T. Hattori, T. Hasuike and S. Wakaizumi, Phys. Rev. {\bf
D60} (1999) 113008.
\bibitem{eta}S. Herrich and U. Nierste, hep-ph/9604330; hep-ph/9310311; 
S. Herrich, hep-ph/9609376.
\bibitem{sxy}J.F. Donoghue, E. Golowich and B.R. Holstein, {\it
Dynamics of the Standard Model} (Cambridge University Press, New York, 1992).
\bibitem{al}M.S. Alam, Phys. Rev. Lett. {\bf 74}, (1995) 2885.
\bibitem{Ali}A. Ali, hep-ph/9606324; hep-ph/9612262.
\bibitem{9905397}G. Barenboim, G. Eyal and Y. Nir, hep-ph/9905397.
\bibitem{hierarchy}M. Leurer, Y. Nir, N. Seiberg, Nucl. Phys. 
{\bf B420} (1994) 468; P. Kielanowski, {\it et al}, hep-ph/0002062.
\end{thebibliography}
\end{document}